# Mechanical, optoelectronic and transport properties of single-layer $Ca_2N$ and $Sr_2N$ electrides

Bohayra Mortazavi[*,1], Golibjon R Berdiyorov[2], Masoud Shahrokhi[3], Timon Rabczuk[#,4]

[1]*Institute of Structural Mechanics, Bauhaus-Universität Weimar, Marienstr. 15, D-99423 Weimar, Germany.*

[2]*Qatar Environment and Energy Research Institute, Hamad Bin Khalifa University, Qatar Foundation, Doha, Qatar.*

[3]*Institute of Chemical Research of Catalonia, ICIQ, The Barcelona Institute of Science and Technology, Av. Països Catalans 16, ES-43007 Tarragona, Spain.*

[4]*College of Civil Engineering, Department of Geotechnical Engineering, Tongji University, Shanghai, China.*

**Abstract**

Electride materials offer attractive physical properties due to their loosely bound electrons. $Ca_2N$, an electride in the two-dimensional (2D) form was successfully recently synthesized. We conducted extensive first-principles calculations to explore the mechanical, electronic, optical and transport response of single-layer and free-standing $Ca_2N$ and $Sr_2N$ electrides to external strain. We show that $Ca_2N$ and $Sr_2N$ sheets present isotropic elastic properties with positive Poisson's ratios, however, they yield around 50% higher tensile strength along the zigzag direction as compared with armchair. We also showed that the strain has negligible effect on the conductivity of the materials; the current in the system reduces by less than 32% for the structure under ultimate uniaxial strain along the armchair direction. Compressive strain always increases the electronic transport in the systems due to stronger overlap of the atomic orbitals. Our results show that the optical spectra are anisotropic for light polarization parallel and perpendicular to the plane. Interband transition contributions along in-plane polarization are not negligible, by considering this effect the optical properties of $Ca_2N$ and $Sr_2N$ sheets in the low frequency regime significantly changed. The insight provided by this study can be useful for the future application of $Ca_2N$ and $Sr_2N$ in nanodevices.

*Keywords:* $Ca_2N$ and $Sr_2N$; 2D electride; mechanical; electronic; optical

*Corresponding author (Bohayra Mortazavi): bohayra.mortazavi@gmail.com

Tel: +49 157 8037 8770; Fax: +49 364 358 4511; [#]Timon.rabczuk@uni-weimar.de

## 1. Introduction

Two-dimensional (2D) materials are considered as the new class of materials for numerous and diverse applications ranging from nanoelectronics to aerospace structures. The successful mechanical exfoliation of graphene from graphite in 2004 [1,2], raised an ongoing interest toward the synthesis of other high quality and large-area 2D materials. In this way, during the last decade a wide variety of 2D materials with outstanding properties have been synthesized such as: hexagonal boron-nitride [3,4], graphitic carbon nitride [5,6], silicene [7,8], germanene [9], transition metal dichalcogenides [10–12], phosphorene [13,14] and most recently borophene [15].

Among the various groups of 2D materials, recently 2D electrides have attracted remarkable attentions. Almost every ionic solid consists of positively and negatively charged atoms, but in electrides, the negative "ion" is just an electron with no nucleus [16,17]. 2D electrides present $M_2N$ (M = Ca, Sr), in which conduction electrons are confined between the M layers with a concentration corresponding to the chemical formula $[M_2N]^+e^-$ [17]. Because of the existence of nucleus-free 2D electron gas in free space, and presence of very high electron mobility along with the low work function, 2D electride materials can be promising candidates for numerous application such as the next-generation electronics. Recent theoretical investigation [18] suggests that the 2D $Ca_2N$ meets all the requirements as a suitable anode for rechargeable Na-ion batteries. Despite of the attractive properties of 2D electrides, exfoliation of their atomic layers to form 2D nanomaterials has been a challenging issue. In the most recent experimental study by Druffel *et al.* [16], for the first time $Ca_2N$ nanomembranes were fabricated. This experimental advance along with the outstanding physical properties of 2D electrides consequently raise the importance of theoretical and experimental studies to explore their properties. The in-depth understanding of 2D electride properties not only plays crucial role for the nanodevices design, but it may also suggest new applications for their usage. Since experimental studies for the evaluation of the properties of 2D materials are complicated, time consuming and expensive, theoretical approaches can be considered as viable alternatives to explore their various properties [19–29]. In the present investigation, we therefore studied the mechanical, electronic and optical properties of single-layer and free-standing $Ca_2N$ and $Sr_2N$ sheet using extensive first-principles density functional theory (DFT) simulations. We first conducted uniaxial tensile and compression simulations to evaluate the key mechanical properties along the both armchair and zigzag directions. We then explored the effect of external strain on the electronic transport properties of both systems using DFT calculations in combination with the Green's functional formalism. In addition, we studied the optical properties and their change under the various loading conditions. The obtained results by the present investigation provide a general vision concerning the mechanical, electronic and optical properties of 2D electrides and therefore can be very useful for their practical applications.

2. Methods

The density functional theory calculations in this study were performed using the Vienna ab-initio simulation package (VASP) [30–32]. We employed the plane wave basis set with an energy cut-off of 500 eV and the gradient approximation exchange-correlation functional proposed by Perdew-Burke-Ernzerhof [33]. We also used VMD [34] and VESTA [35] packages for the illustration of the structures. Fig.1, illustrates the hexagonal lattice of $Ca_2N$ or $Sr_2N$ atomic structure which shows ABC atomic stacking sequence of considered structures. In this work we studied the properties upon the stretching along the chirality directions of armchair and zigzag, as also shown in Fig. 1. For the stretching along the armchair and zigzag, we constructed cubic supercells with 24 and 18 atoms, respectively. We applied periodic boundary conditions along all three Cartesian directions and such that the obtained results represent the properties of large-area single-layer films and not the nanoribbons. We should however note that since the dynamical effects such as the temperature are not taken into consideration in the DFT modelling and periodic boundary conditions were also applied along the planar directions, only a unit-cell modelling [36] with 6 atoms was accurate enough to investigate the mechanical properties of 2D electrides. To avoid image-image interactions along the sheet normal direction, we considered a vacuum layer of 20 Å in this direction [37]. After obtaining the minimized structure, we applied loading conditions to evaluate the mechanical properties. For this purpose, we increased the periodic simulation box size along the loading direction in a multistep procedure, every step with a small engineering strain of 0.001. After applying the changes in the simulation box size, the atomic positions were rescaled to avoid any sudden void formation or bond stretching as well. We then used the conjugate gradient method for the geometry optimizations, with strict termination criteria of $10^{-5}$ eV and 0.01 eV/Å for the energy and the forces, respectively, using a 6×6×1 Monkhorst-Pack [38] $k$-point mesh size. The final stress values after the termination of energy minimization process were calculated to obtain the stress-strain curves.

Structural parameters from the VASP calculations were then used to calculate the electronic transport properties of $Ca_2N$ and $Sr_2N$ using the computational package Atomistix toolkit. The electronic structure calculations were conducted using the DFT/PBE and the electrostatic potentials were obtained on a real-space grid with a mesh cutoff energy of 150 Ry using the nonequilibrium Green's function formalism. All the atoms are described using double-zeta-polarized basis of numerical orbitals. The current-voltage (*I-V*) characteristics are calculated using the Landauer-Buttiker formula:

$$I(V) = \frac{2e}{h} \int_{\mu_L}^{\mu_R} T(E,V)[f(E-\mu_L) - f(E-\mu_R)] dE \qquad (1)$$

where *T(E,V)* is the transmission spectrum for the given value of voltage biasing (V), $f(E,E_f)$ is the Fermi-Dirac distribution function and $\mu_L/\mu_R$ is the chemical potential of the left/right electrode. In order to calculate the electronic transport, we constructed device geometries for each strain levels, (see Figs. 2 for the case of unstrained $Ca_2N$). They consist of left and right electrode regions and a central (scattering) region (i.e. two-probe configuration). 100 *k*-points are used along the transport directions and 12 *k*-point sampling was performed in the other planar direction.

Optical calculations in this paper were performed in the random phase approximation (RPA) using all electron, full potential Wien2k [39] code. Maximum angular momentum of the atomic orbital basis functions was set to $l_{max} = 10$. In order to achieve energy eigenvalues convergence, wave functional in the interstitial region were expanded in terms of plane waves with a cut-off parameter of $RMT \times K_{max} = 8.5$, where RMT denotes the smallest atomic sphere radius and $K_{max}$ largest *k* vector in the plane wave expansion. Magnitude of largest vector in charge density Fourier expansion or the plane wave cutoff were set to $G_{max} = 12$. The optical spectra were calculated using 24×24×1 Γ centered Monkhorst-Pack [38] *k*-point mesh in the first Brillouin zone and setting Lorentzian broadening with gamma equal to 0.05 eV.

The optical properties are determined by the frequency dependent dielectric function $\varepsilon_{\alpha\beta}(\omega) = \text{Re}\,\varepsilon_{\alpha\beta}(\omega) + i\,\text{Im}\,\varepsilon_{\alpha\beta}(\omega)$, which mainly depends on the electronic structure. The imaginary part $\text{Im}\,\varepsilon_{\alpha\beta}(\omega)$ of the dielectric function by considering interband transitions is obtained using of following relation [40,41]:

$$\text{Im}\,\varepsilon_{\alpha\beta}^{(\text{inter})}(\omega) = \frac{4\pi^2 e^2}{\Omega}\lim_{q\to 0}\frac{1}{|q|^2}\sum_{c,v,k}2w_k\delta(\varepsilon_{ck}-\varepsilon_{vk}-\omega)\times\langle u_{ck+e_\alpha q}|u_{vk}\rangle\langle u_{ck+e_\beta q}|u_{vk}\rangle^* \qquad (2)$$

where *q* is the Bloch vector of the incident wave and $w_k$ the **k**-point weight. The band indices *c* and *v* are restricted to the conduction and the valence band states, respectively. The vectors $e_\alpha$ are the unit vectors for the three Cartesian directions, $\Omega$ is the volume of the unit cell and $u_{ck}$ is the cell periodic part of the orbitals at the $k$-point **k**. The real part $\text{Re}\,\varepsilon_{\alpha\beta}(\omega)$ is obtained by the Kramers–Kronig transformation from its corresponding $\text{Im}\,\varepsilon_{\alpha\beta}(\omega)$ [42]:

$$\text{Re}\,\varepsilon_{\alpha\beta}^{(\text{inter})}(\omega) = 1 + \frac{2}{\pi} P \int_0^\infty \frac{\omega' \text{Im}\,\varepsilon_{\alpha\beta}(\omega')}{(\omega')^2 - \omega^2 + i\eta} d\omega' \quad (3)$$

In this equation $P$ denotes the principle value and $\eta$ is the complex shift. Because of metallic properties of these nanosheets, the intraband transitions contribution should be added to the interband transitions to investigate the optical properties of these systems. By taking into account the contribution of intraband transitions for metals, it is obtained [29]:

$$\text{Im}\,\varepsilon_{\alpha\beta}^{[\text{intra}]}(\omega) = \frac{\Gamma \omega_{pl,\alpha\beta}^2}{\omega(\omega^2 + \Gamma^2)} \quad (4)$$

$$\text{Re}\,\varepsilon_{\alpha\beta}^{[\text{intra}]}(\omega) = 1 - \frac{\omega_{pl,\alpha\beta}^2}{\omega(\omega^2 + \Gamma^2)} \quad (5)$$

where $\omega_{pl}$ is the plasma frequency and $\Gamma$ is the lifetime broadening.

### 3. Results and discussions

The atomic structure of $Ca_2N$ and $Sr_2N$ can be defined by the hexagonal lattice constant ($\alpha$) and Ca–N or Sr–N bond length. Using the DFT calculations for the cell energy minimization and geometry optimization, the lattice constant of $Ca_2N$ and $Sr_2N$ were calculated to be 3.62 Å and 3.87 Å, respectively. The Ca–N and Sr–N bond lengths were also obtained to be 2.44 Å and 2.62 Å, respectively. The lattice parameters we obtained in this work are maximum in less than 0.4% difference with those reported in an earlier study [18]. We note that in the previous investigation [43] the dynamical stability of $Ca_2N$ and $Sr_2N$ was confirmed by calculating the phonon dispersion relations in which no negative frequency was observed. To examine the energetic stability of the 2D $Ca_2N$ and $Sr_2N$, the cohesive energy per atom was calculated as defined by:

$$E_{coh} = (\sum_i E_i - E_t)/n \qquad (6)$$

where $E_t$, $E_i$ and $n$ denote the total energy per cell, the energy of the $i$-th isolated atom and the total number of atoms in the cell, respectively. Our results reveal that the cohesive energy for these structures are negative, -4.59 for $Ca_2N$ and -4.21 for $Sr_2N$, which confirms that these structures are energetically stable. To probe the electronic properties of strain-free $Ca_2N$ and $Sr_2N$ nanostructure, the band structure along the high symmetry $\Gamma$-$M$-$K$-$\Gamma$ directions, and total density of states (DOS) were calculated using the PBE approach and the obtained results are illustrated in Fig. 3. The calculations show that the Fermi level intersects band structures and DOS in both compounds which accordingly demonstrate a metallic characteristic and are in excellent agreements with previous work [43].

To evaluate the elastic properties of single-layer $Ca_2N$ and $Sr_2N$, we first applied uniaxial strains. In this case, we applied the strain along the loading direction while the simulation box size along the transverse direction was fixed (strain in transverse direction was zero). In Fig. 4, the DFT prediction for stress-strain responses of $Ca_2N$ and $Sr_2N$ elongated along the armchair direction are plotted. Here, the stress values along the loading ($\sigma_l$) and transverse ($\sigma_t$) directions are plotted at each strain level. It is seen from this figure that up to the strain level of ~0.02, the results for the both stress components present linear relations which reveals that the specimens are stretched within their elastic regime. Using the obtained stress-strain curves the elastic modulus and Poisson's ratio can be calculated based on the Hooke's Law. The Hooke's Law for a 2D material with orthotropic elastic properties, can be written as:

$$\begin{bmatrix} s_{xx} \\ s_{yy} \end{bmatrix} = \begin{bmatrix} \dfrac{1}{E_x} & \dfrac{-\nu_{yx}}{E_y} \\ \dfrac{-\nu_{xy}}{E_x} & \dfrac{1}{E_y} \end{bmatrix} \begin{bmatrix} \sigma_{xx} \\ \sigma_{yy} \end{bmatrix} \quad (7)$$

here $s_{ii}$, $\sigma_{ii}$, $v_{ij}$ and $E_i$ are the strain, stress, Poisson's ratio and elastic modulus along the "$i$" direction, respectively. Since we only applied a uniaxial strain along the "$x$" direction, the strain is zero along the transverse direction ($s_{yy}=0$), and such that:

$$v_{xy} = \frac{\sigma_{yy}E_x}{\sigma_{xx}E_y} \quad (8)$$

However, based on the symmetry of the stress and strain tensors, the following relation exist:

$$\frac{v_{yx}}{v_{xy}} = \frac{E_y}{E_x} \quad (9)$$

By replacing the Eq. 9 in Eq. 8, one can obtain:

$$v_{yx} = \frac{\sigma_{yy}}{\sigma_{xx}} \quad (10)$$

We remind that in our loading condition, $\sigma_{xx}$ and $\sigma_{yy}$ are equivalent with $\sigma_l$ and $\sigma_t$, respectively. By applying the uniaxial strains along the armchair and zigzag directions, the Poisson's ratio along the both loading directions can be obtained using the calculated stress values. In the next step, using the Eq. 7, the elastic modulus can be calculated along the both armchair and zigzag directions. Based on our DFT modelling, we found that the elastic modulus and Poisson's ratio of $Ca_2N$ and $Sr_2N$ membranes along the armchair and zigzag directions are the same and this reveal isotropic elastic properties for the both studied 2D materials. For single-layer $Ca_2N$ and $Sr_2N$, the elastic modulus are predicted to be 56 GPa.nm and 43 GPa.nm, respectively. The Poisson's ratios of 2D $Ca_2N$ and $Sr_2N$ are also calculated to be 0.26 and 0.29, respectively.

We then study the mechanical response $Ca_2N$ and $Sr_2N$ by performing the uniaxial tensile simulations. For the uniaxial loading conditions, upon the stretching along the loading direction the stress along the transverse direction should be negligible. To satisfy this condition, after applying the loading strain, the simulation box size along the transverse direction of the loading was changed accordingly in a way that the transverse stress remained negligible in comparison with the stress along the loading direction. The DFT predictions for the uniaxial stress-strain responses of pristine and free-standing $Ca_2N$ and $Sr_2N$ stretched along the armchair and zigzag loading directions are illustrated in Fig. 5. In all cases, the stress-strain response presents an initial linear relation which is followed by a nonlinear curve up to the ultimate tensile strength point, a point at which the material illustrates its maximum load bearing. Because of the isotropic elastic response of the both studied electrides, the initial linear region of the stress-strain curves coincide for the stretching along the armchair and zigzag directions. However, our DFT results reveal that the tensile response is not isotropic and along the zigzag direction the single-layer $Ca_2N$ and $Sr_2N$ are by ~50% stronger as compared with armchair. In addition, along the zigzag direction the structures are around twice more stretchable in comparison with the armchair direction. For the $Ca_2N$, the tensile strength along the zigzag and armchair are predicted to be ~6.5 GPa.nm and ~4.4 GPa.nm, respectively. In this case, the strain at tensile strength for the uniaxial loading along zigzag and armchair are ~0.24 and ~0.14, respectively. For the $Sr_2N$, we also found similar trends, however, the tensile strength and its corresponding strain are lower than $Ca_2N$. The tensile strength of $Sr_2N$ elongated along the zigzag and armchair direction are obtained to be ~4.7 GPa.nm and ~3.1 GPa.nm, respectively.

To better understand the underlying mechanism that results in the anisotropic tensile response of $Ca_2N$ and $Sr_2N$, we analyzed the deformation process. Since the deformation behaviour is similar for $Ca_2N$ and $Sr_2N$, we here only study the $Ca_2N$ membrane. Fig. 6, compares the deformation process of free-standing $Ca_2N$ elongated along the armchair and zigzag directions. We selected 3 samples at different strain levels with respect to the strain at the tensile strength point, $s_{uts}$. As a general observation, during the uniaxial tensile loading, the bonds that were oriented along the loading direction were elongated and on the other hand those bonds that were oriented almost along the transverse direction were contracted slightly. For the stretching along the armchair direction, from every three bonds, two Ca–N bonds are oriented almost along the transverse direction. In this case, only one Ca–N bond, which is exactly along the loading direction is incorporated in the load bearing. Therefore, along the armchair direction the stretchability of this bond plays crucial in the mechanical response. Along the zigzag, in the unit-cell two Ca–N bonds are almost oriented along the loading direction and the other bond is exactly along the transverse direction. This way, for the $Ca_2N$ nanomembranes stretched along the zigzag more bonds are involved in stretching and accordingly in the load transfer. We found that at the tensile strength point, the simulation box size along the transverse direction for the $Ca_2N$ elongated along the armchair and zigzag, shrinks by ~0.3% and ~0.6%, respectively. As it is shown in Fig. 6, along the both loading directions, during the stretching the $Ca_2N$ films contract slightly along the sheet thickness. In this case, the decrease in the sheet thickness is more considerable for the sample elongated along the zigzag direction. The higher stretchability along the zigzag can be therefore attributed to the fact that the structure can contract more in the transverse and sheet thickness directions before reaching to the critical bond length at the tensile strength point. On the other side, the higher tensile strength along the zigzag direction in comparison with the armchair can be explained by the fact that more bonds are involved in the stretching and consequently load bearing during the uniaxial tensile loading along the zigzag.

Next, we study the effect of external strain (both compressive and tensile) on the electronic transport properties of the considered 2D materials. As a main result, we present in Fig. 7, the *I-V* curves of (a) $Ca_2N$ and (b) $Sr_2N$ for different values of uniaxial loading and for the bias voltage below 1 V. The electronic transport is calculated along the armchair direction. In their unstrained form, both samples show similar *I-V* curves with maximal current around 30 $\mu A$ (filled-black circles). In both cases, compressive strain increases the conductance for the considered range of voltage biasing (open-red circles). For the same level, of compression, the current enhancement is more pronounced in the case of $Ca_2N$. This is shown in the insets of Fig. 7, where we plot the current (normalized to the one in the unstrained sample) as a function of strain for two values of the voltage biasing, $\Delta V$. Larger increase in the current due to the compression is obtained for larger $\Delta V$ (open-red circles in the insets of Fig. 7). For the considered range of the compressive strain and applied voltage, the current in the system can be increased more than 10%. On the contrary, tensile strain decreases the current in the system, where the level of decrease in the current shows (quasi)linear dependence on the tensile loading (see the insets of Fig. 7). However, the maximum decrease in the conduction not exceed 32 % for the ultimate strain (see open triangles in Fig. 7).

To understand the origin for the change in the electronic transport, we have calculated the electron difference density (EDD) in the considered samples, which gives the difference between the self-consistent valence charge density and the superposition of atomic valence densities. Fig. 8 shows the EDD of $Ca_2N$ for different strain values. In unstrained sample the electronic states are delocalized along both planar directions (Fig.8b). Compressive strain (Fig. 8a) increases the overlap of the electronic orbitals, which ultimately results in the enhanced electronic transport. On the contrary, tensile strain results in localization of electronic states (Fig. 8). Such reduced overlap of electronic states along the transport direction reduces the probability of electrons to cross the system. Thus, the localization of electronic states plays a key role in determining the transport properties of the considered systems. Note that we obtained similar results for the $Sr_2N$ sample.

Variations of the electrostatic potential is also an important factor affecting the electronic transport. To see the strain effect on the potential variations, we present in Fig. 9 the averaged electrostatic difference potential in $Ca_2N$ along the transport direction for different strain levels at zero voltage biasing. Periodic potential oscillations are obtained for the unstrained sample (solid-black curve in Fig. 9). Compressive strain decreases the amplitude of the potential oscillations (dashed-red curve in Fig. 9), which reduces the scattering of the electrons from the active region of our device. On the contrary, tensile strain increases the amplitude of the oscillations (see dotted-green and dash-dotted-blue curves in Fig. 9). Electron scattering from these potential oscillations reduce the conductivity of the system.

Finally, we study the optical properties of the considered materials under different external strain conditions. Fig. 10 (a-d) illustrates Im $\varepsilon$ and Re $\varepsilon$ of strained $Ca_2N$ monolayer sheet along the armchair (a and b) and zigzag (c and d) direction at different strain levels for light polarizations parallel (E||x and E||y) and perpendicular (E||z) to the plane (up to 10 eV). The top panel of each figure illustrates the optical spectra considering the interband transitions (inter) contribution. The lower panel of these figures show the optical spectra where both interband + intraband (inter + intra) transitions contributions are taken into account. It can be seen that for all strained and unstrained $Ca_2N$ monolayer systems Im $\varepsilon$ obtained from interband transitions contribution in both polarizations starts without a gap, confirms that all aforementioned systems have metallic property. By adding the intraband transitions contribution, the optical spectra of these systems in the electric field parallel to plane (for both E||x and E||y) in the low-frequency regime (up to 2 eV) is changed significantly, while it remains unchanged in the electric field perpendicular to plane E||z which shows the weak metallic properties for perpendicular polarization of these systems. Moreover, it is found that by taking into account the intraband transitions contribution, the Re $\varepsilon$ for all cases in E||x and E||y is changed significantly, while it does not have any effect on the Re $\varepsilon$ in perpendicular polarization.

Similar optical response to the external strain has been obtained for $Sr_2N$ monolayer. This is shown in Fig. 11 (a-d) where we plot Im $\varepsilon$ and Re $\varepsilon$ of strained $Sr_2N$ monolayer sheet along the armchair (a and b) and zigzag (c and d) direction at different magnitudes of strains for parallel and perpendicular light polarizations. It is known that the roots of $\text{Re}\,\varepsilon_{\alpha\beta}(\omega)$ with x = 0 line show the plasma frequencies [29]. Table 1 summarizes our results of the first plasma frequency for $Ca_2N$ and $Sr_2N$ monolayer sheets at different magnitudes of strain along $E\|x$, $E\|y$ and $E\|z$. The first plasma frequency for all systems in $E\|z$ is nearly zero indicating metallic nature of the materials properties in perpendicular polarization. This can be justified by the huge depolarization effect along $E\|z$ [44] which leads to same results for graphene [45] and other metallic 2D sheets like $B_2C$ monolayer [46]. In general, the imaginary and real parts of dielectric function for all systems have singularity at zero frequency in parrallel polarizations after adding the intraband transitions contribution because of the strong metallic behavior along these polarizations. Forthermore, for all aforementioned systems the dielectric function becomes anisotropic along armchair, zigzag and out-of plane directions.

## 4. Conclusion

We conducted extensive first-principles DFT calculations to explore the mechanical, optical and transport response of pristine and single-layer $Ca_2N$ and $Sr_2N$ electrides. We showed that $Ca_2N$ and $Sr_2N$ exhibit isotropic elastic properties, and their elastic modulus are predicted to be 56 GPa.nm and 43 GPa.nm, respectively. The Poisson's ratios of 2D $Ca_2N$ and $Sr_2N$ are also calculated to be ~0.26 and ~0.29, respectively. Based on our DFT modelling their mechanical responses however were found to be anisotropic in which they yield around 50% higher tensile strength along the zigzag direction as compared with armchair. This was explained by the fact that for the sample stretched along the zigzag direction more bonds are involved in stretching and accordingly in the load transfer, in comparison with the one elongated along the armchair. We have also conducted DFT calculations in combination with the Green's functional formalism to study the electronic transport properties of free-standing $Ca_2N$ and $Sr_2N$ monolayers under biaxial strain. We found that compressive strain increases the charge transport in the system, whereas tensile loading results in the reduced current. However, the level of reduction in the current does not exceed 32% for the ultimate tensile strain. The change in the conductance originates from the localization of electronic states and electrostatic potential variations in the system. Optical properties of the considered materials are calculated taking into account either only interband transitions or inter and intraband transition. By adding the intraband transition contribution, the optical spectra of relaxed and stretched systems in E∥x and E∥y are changed significantly which indicates strong metallic properties along these polarizations. While the optical spectra in the electric field perpendicular to plane remained unchanged, indicating weak metallic property in E∥z. The plasma frequencies of all systems were reported for both parallel and perpendicular polarizations. Our results also show that the optical spectra of these nanostructures are anisotropic for light polarization parallel and perpendicular to the plane.

## Acknowledgment

B. M. and T. R. greatly acknowledge the financial support by European Research Council for COMBAT project (Grant number 615132). G.R.B. acknowledges computational resources provided by the research computing center at Texas A&M University in Qatar.

Table 1. Calculated the first plasma frequency of $Ca_2N$ and $Sr_2N$ compounds at different magnitudes of strain along the armchair and zigzag directions for both parallel and perpendicular polarizations.

| System | | $E\|\|x$ | | $E\|\|y$ | | $E\|\|z$ | |
|---|---|---|---|---|---|---|---|
| | | Armchair | zigzag | Armchair | zigzag | Armchair | zigzag |
| $Ca_2N$ | s=0 | 1.82 | 1.88 | 1.88 | 1.82 | 0.01 | 0.01 |
| | s=0.25 $s_{uts}$ | 1.46 | 2.05 | 2.01 | 1.83 | 0.03 | 0.03 |
| | s= $s_{uts}$ | 1.89 | 1.99 | 1.95 | 1.81 | 0.03 | 0.05 |
| | s=-0.25 $s_{uts}$ | 2.09 | 2.05 | 1.83 | 1.45 | 0.04 | 0.05 |
| $Sr_2N$ | s=0 | 1.87 | 1.90 | 1.90 | 1.87 | 0.07 | 0.07 |
| | s=0.25 $s_{uts}$ | 1.54 | 2.85 | 1.85 | 1.51 | 0.06 | 0.05 |
| | s= $s_{uts}$ | 1.88 | 1.89 | 1.82 | 1.66 | 0.06 | 0.05 |
| | s=-0.25 $s_{uts}$ | 1.58 | 1.86 | 1.93 | 1.56 | 0.07 | 0.08 |

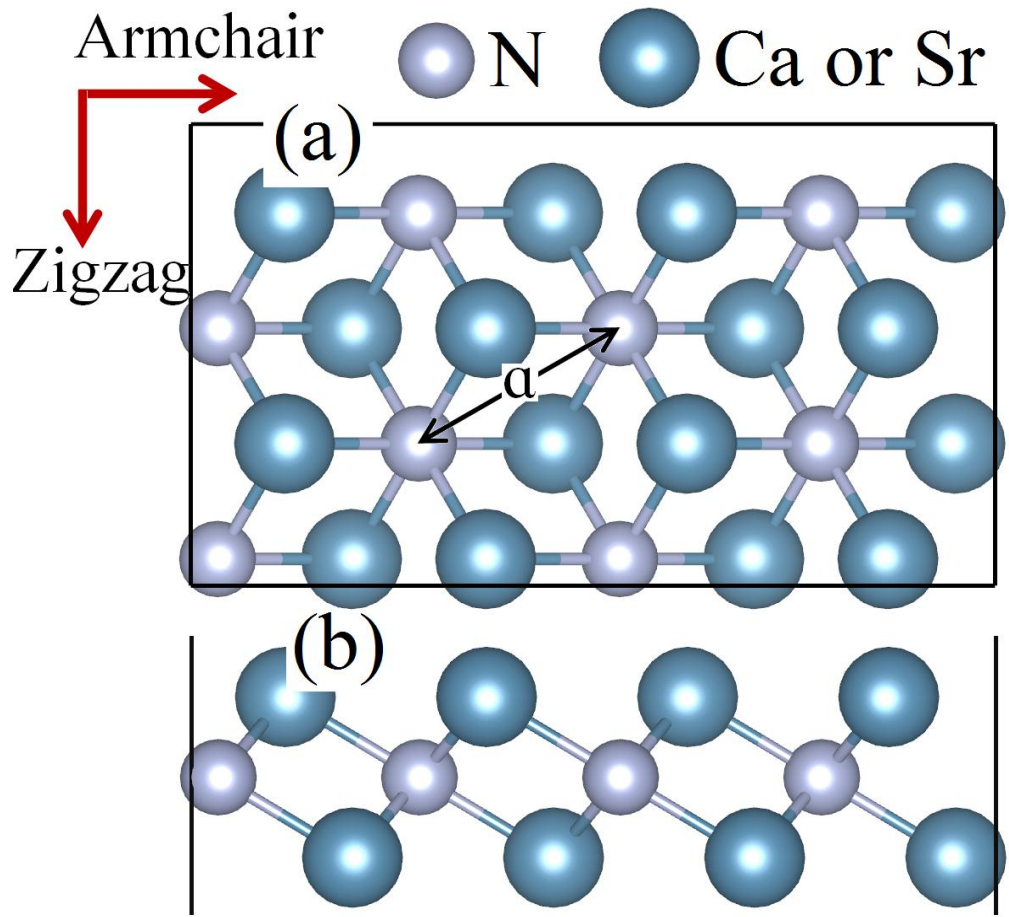

Fig. 1, (a) Top and (b) side views of atomic configuration in $Ca_2N$ or $Sr_2N$ lattice. Mechanical and transport properties are studied along the armchair and zigzag directions as illustrated here.

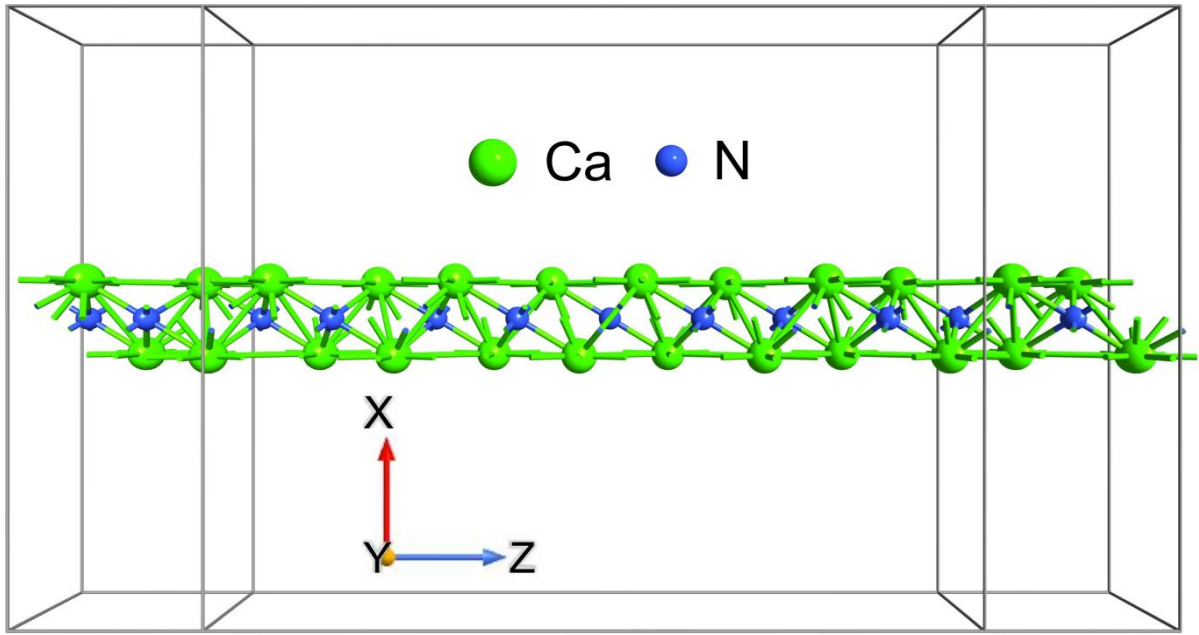

Fig. 2, Device geometries of $Ca_2N$ or $Sr_2N$. The electronic transport is calculated along the armchair direction using 100×12×1 *k*-point sampling.

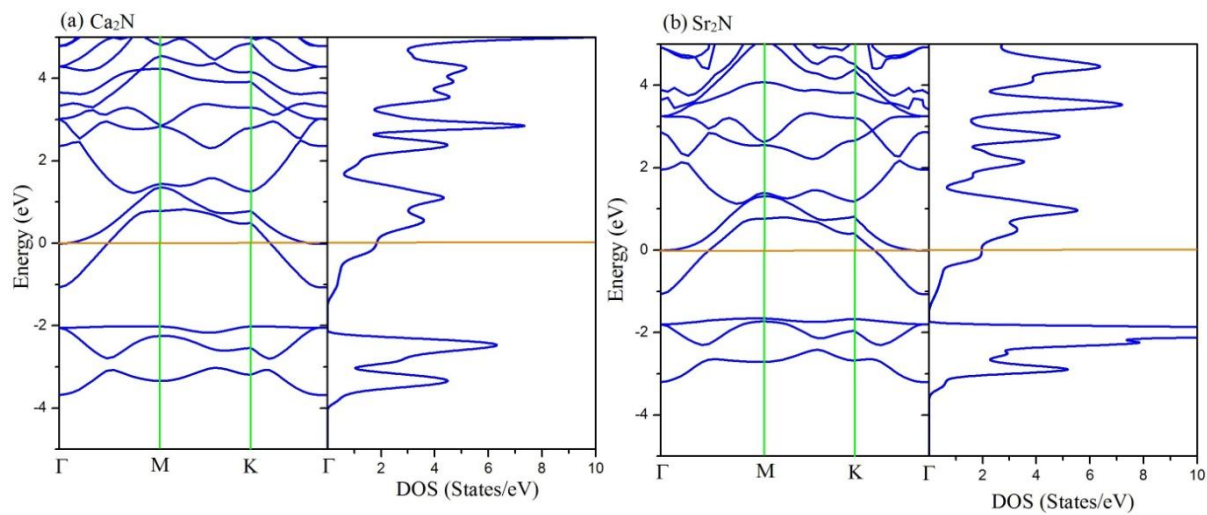

Fig. 3, The band structure and corresponding total DOS predicted for single-layer and strain-free (a) $Ca_2N$ and (b) $Sr_2N$.

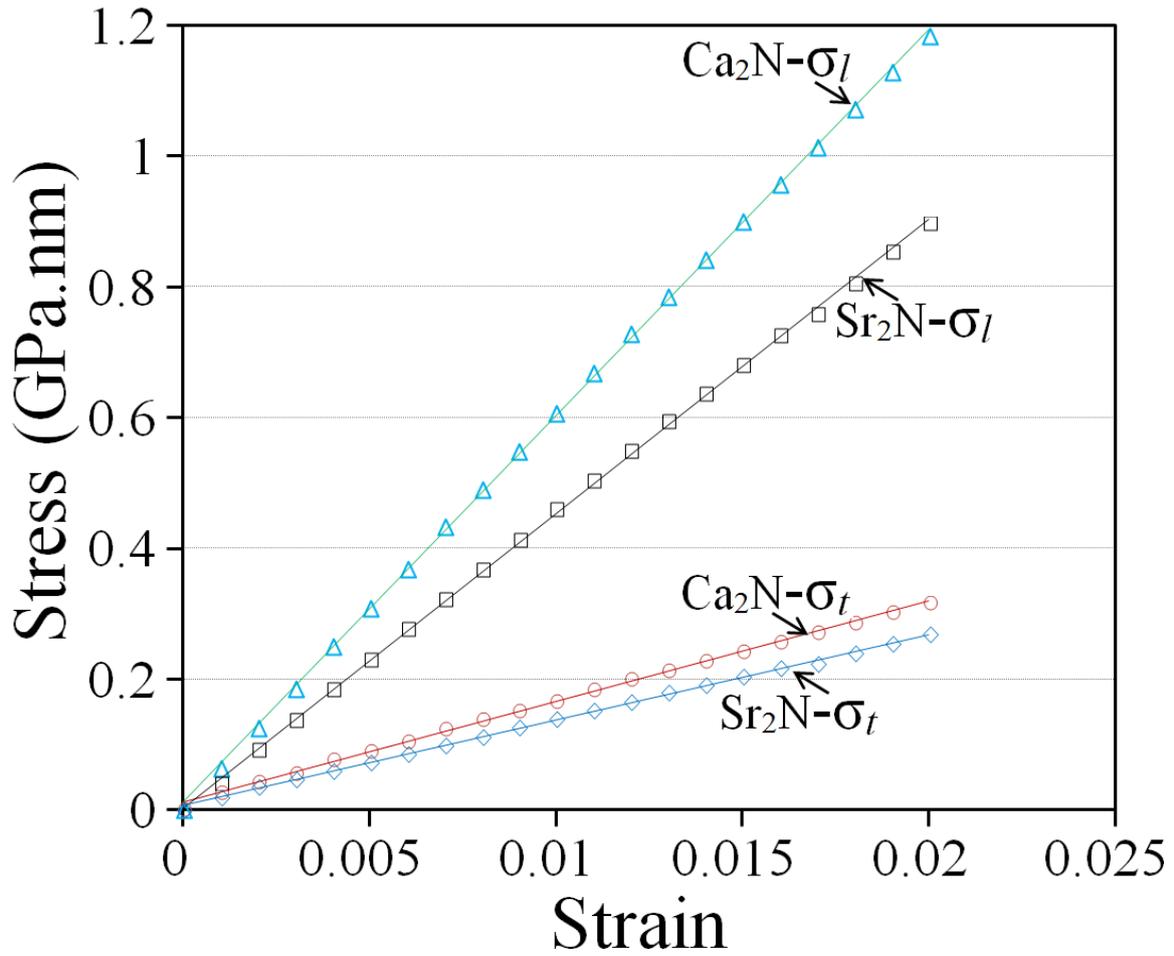

Fig. 4, DFT predictions for the stress-strain responses of single-layer $Ca_2N$ and $Sr_2N$ uniaxially strained along the armchair directions. Here, $\sigma_l$ and $\sigma_t$ denote the stress values along the loading and transverse directions, respectively. Using the Hooke's Law the elastic modulus and Poisson's ratio were then calculated.

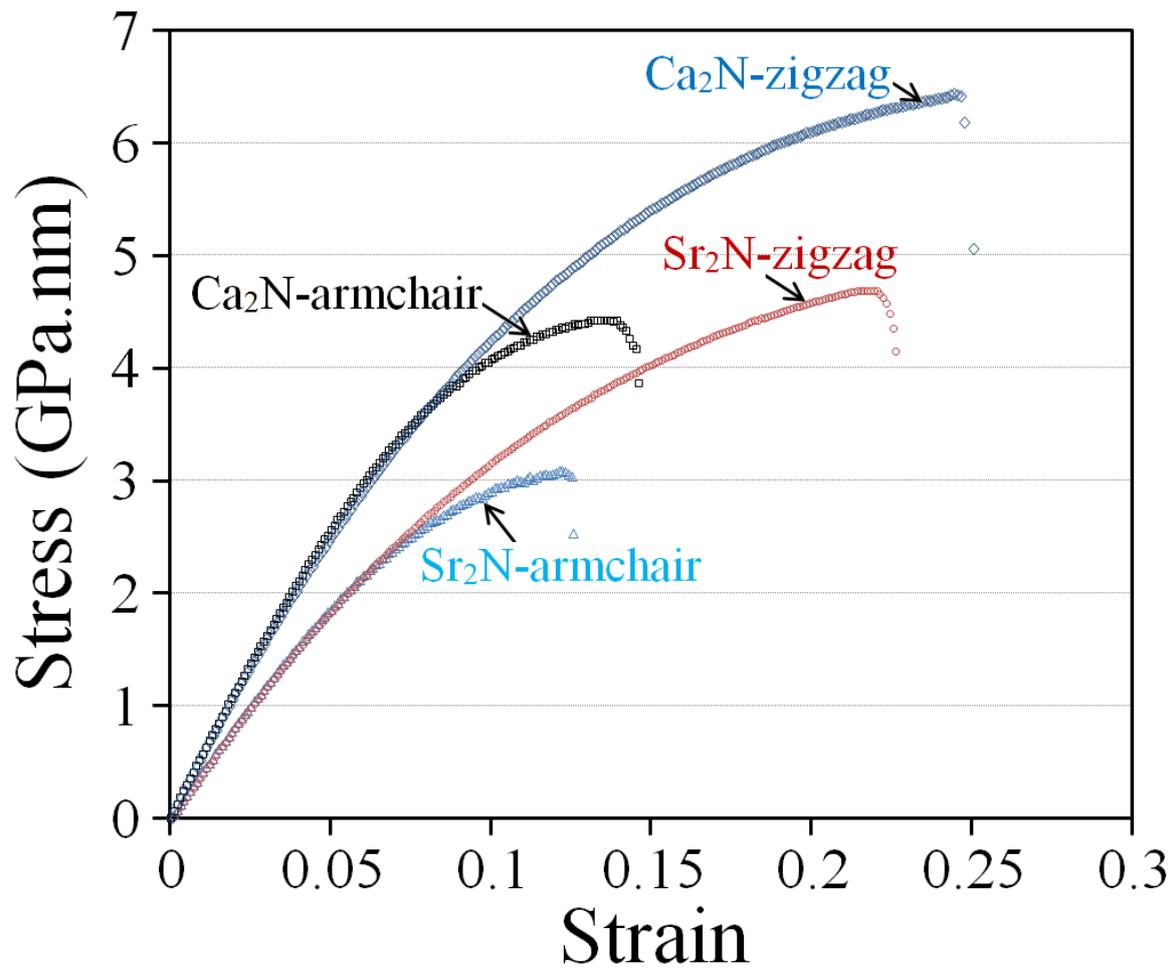

Fig. 5, Uniaxial stress-strain responses of pristine single-layer $Ca_2N$ and $Sr_2N$ stretched along the armchair and zigzag directions.

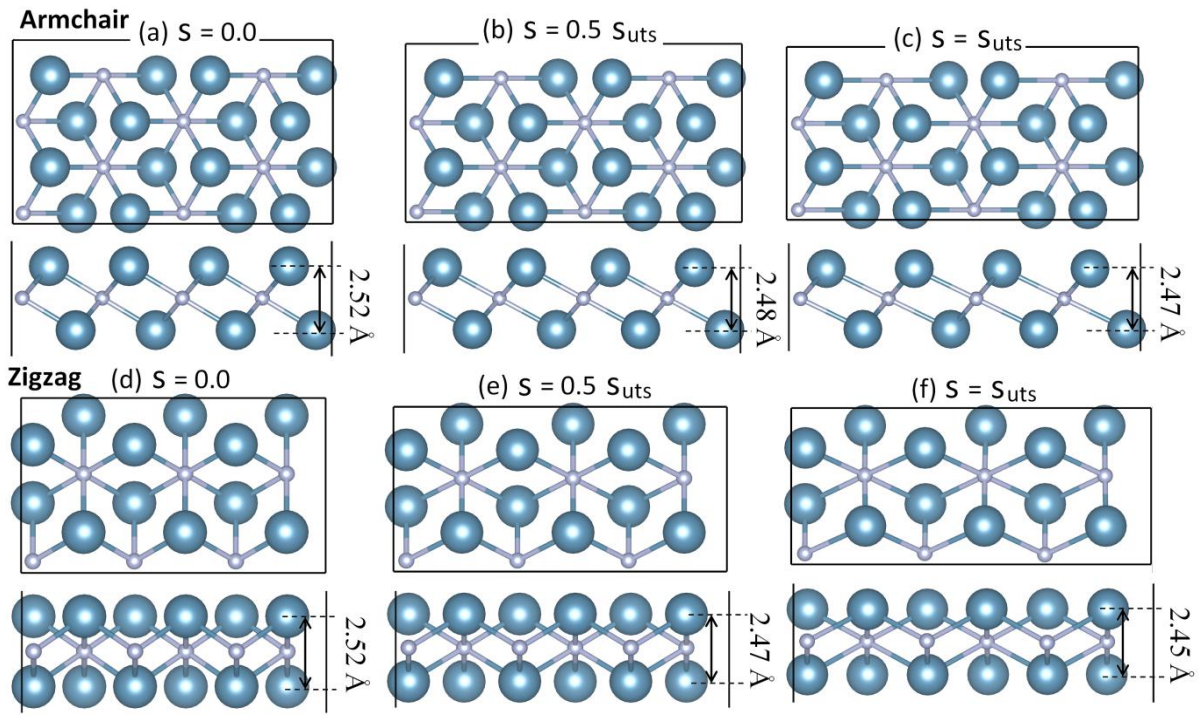

Fig. 6, Top and side views of uniaxial tensile deformation processes of single-layer $Ca_2N$ stretched along the armchair (a-c) and zigzag (d-f) at different strain levels (s) with respect to the strain at ultimate tensile strength ($s_{uts}$).

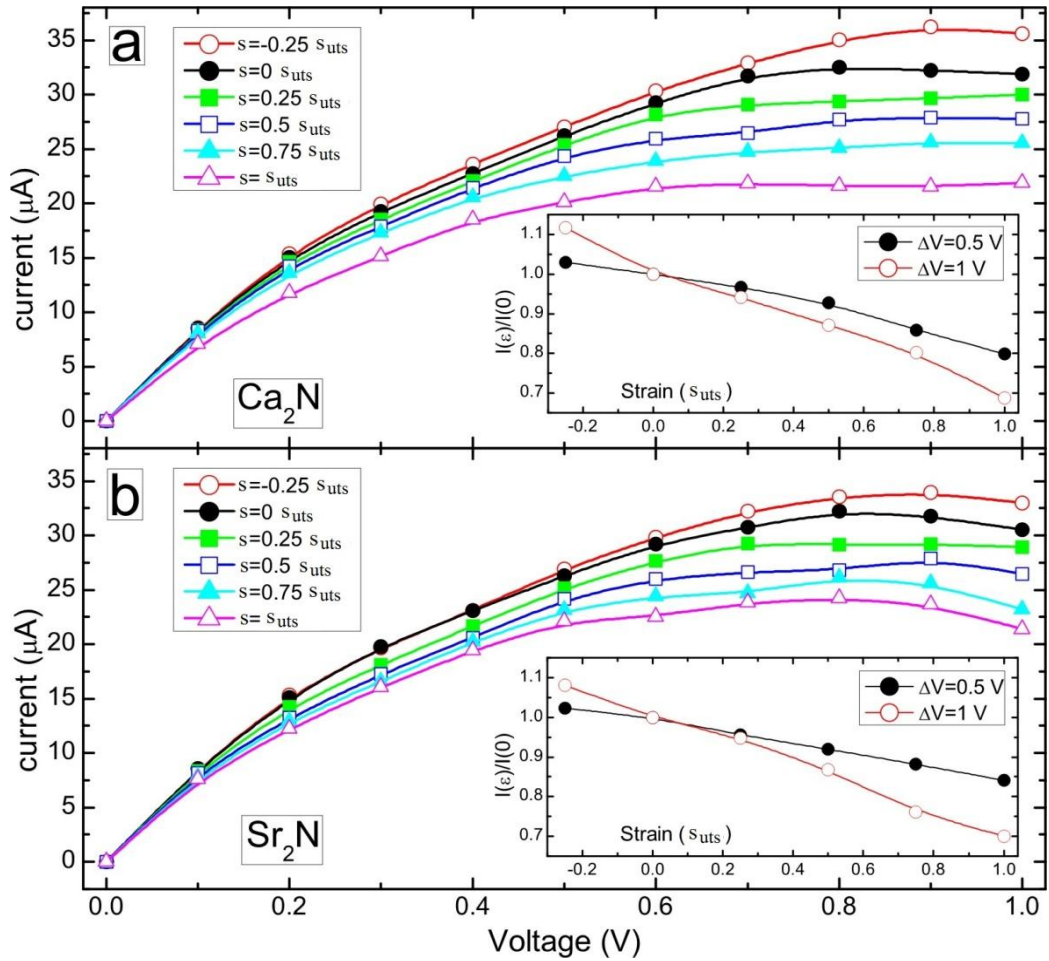

Fig. 7, Current-voltage characteristics of (a) $Ca_2N$ and (b) $Sr_2N$ for different strain levels. The electronic transport is calculated along the armchair direction. Insets show normalized current as a function of external strain at bias voltage $\Delta V = 0.5V$ (filled symbols) and $\Delta V = 1V$ (open symbols). For each case, the results are depicted for different strain levels, s, with respect to the strain at ultimate tensile strength ($s_{uts}$).

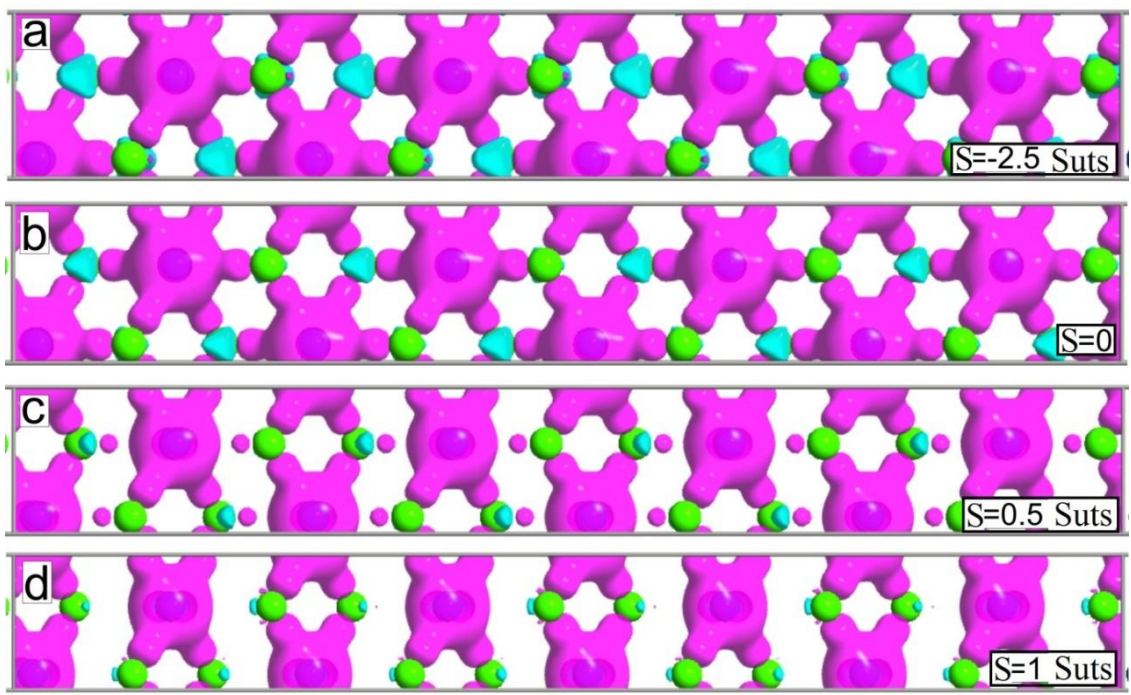

Fig. 8, Isosurface plots (isovalues ±0.04 Å$^{-3}$) of electron difference density for pristine Ca$_2$N under different values of external strain.

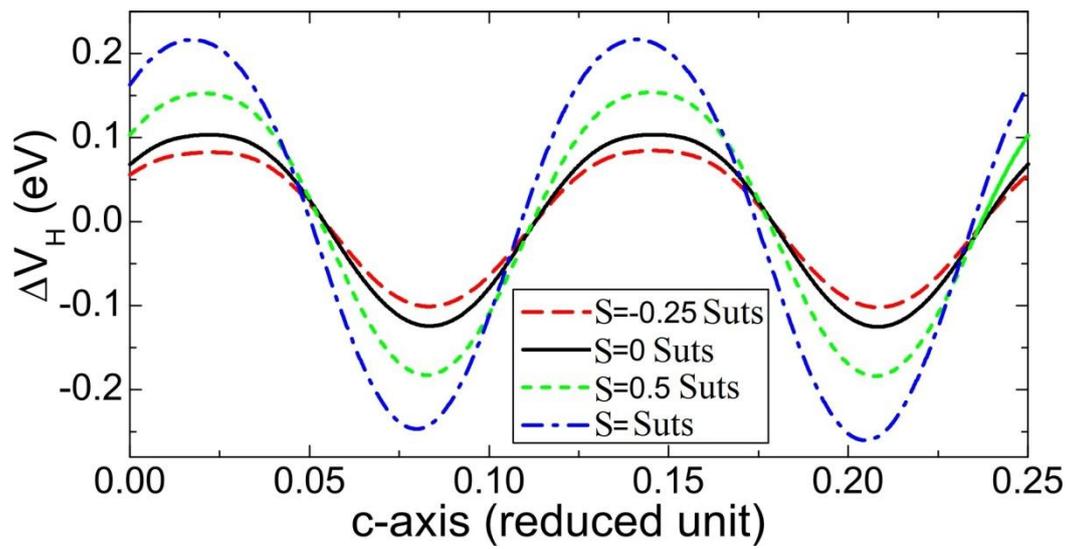

Fig. 9, Averaged electrostatic difference potential along the transport direction for different values of external strain for single-layer $Ca_2N$ stretched along the armchair.

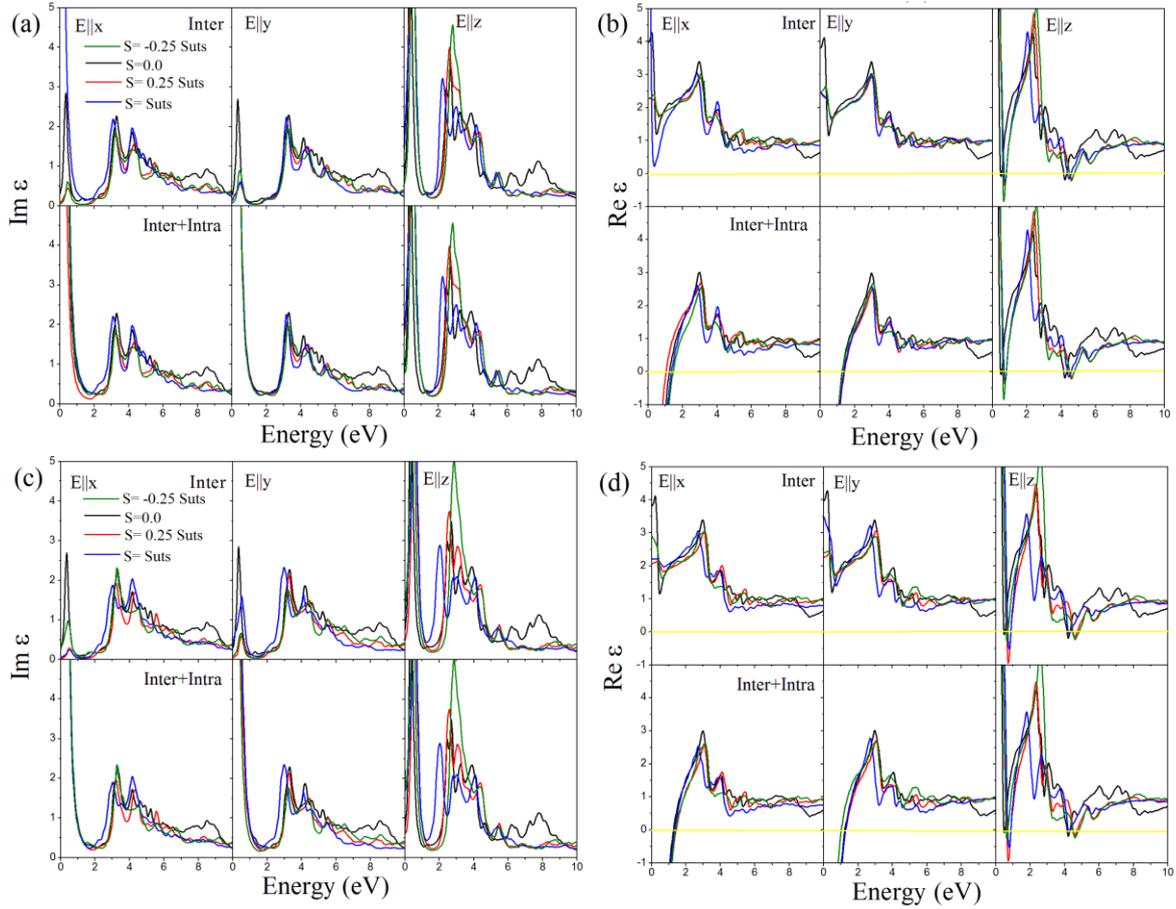

Fig. 10, (a) imaginary and (b) real part of the dielectric function of strained $Ca_2N$ along the armchair direction at different magnitudes of strains for light polarizations parallel (E||x and E||y) and perpendicular (E||z) to the plane. (c) and (d) illustrate the same figures of strained $Ca_2N$ along the zigzag direction at different levels of strain. Below panel in each figure illustrates optical spectra considering the intraband contribution in addition to the interband one.

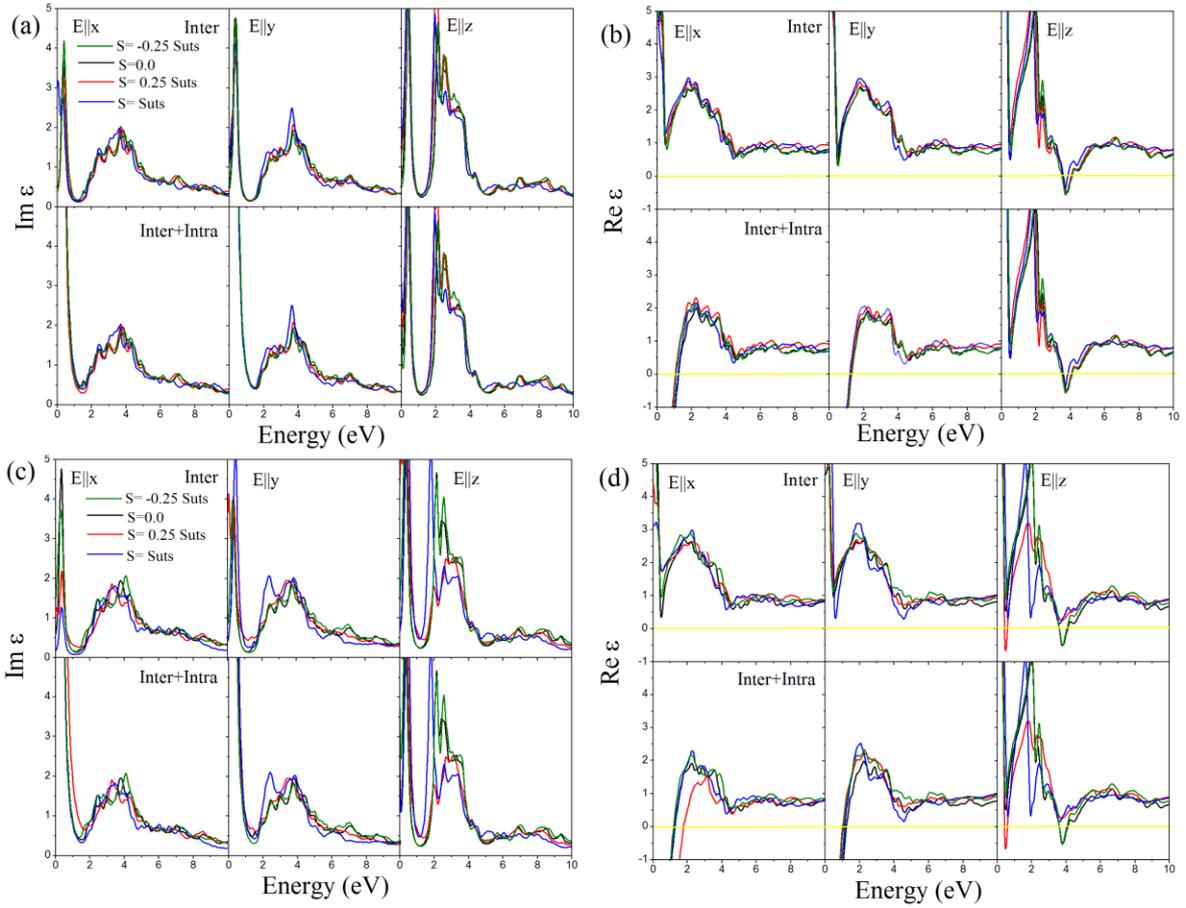

Fig. 11, (a) imaginary and (b) real part of the dielectric function of strained $Sr_2N$ along the armchair direction at different magnitudes of strains for light polarizations parallel (E||x and E||y) and perpendicular (E||z) to the plane. (c) and (d) illustrate the same figures of strained $Sr_2N$ along the zigzag direction at different strain levels. Below panel in each figure illustrates optical spectra considering the intraband contribution in addition to the interband one.